\documentstyle[12pt,a4wide,epsf]{article}

\newcommand{\be}{\begin{equation}}
\newcommand{\ee}{\end{equation}}
\newcommand{\bea}{\begin{eqnarray}}
\newcommand{\eea}{\end{eqnarray}}
\newcommand{\br}{\hskip .25cm/\hskip -.25cm}

\newcommand{\nn}{\nonumber\\}

\newcommand{\ovl}{\overline}

\begin{document}
\begin{flushleft}
KCL-PH-TH/2011-31 \\
LCTS/2011-16 \\
CERN-PH-TH/2011-240
\end{flushleft}
\vspace{0.5cm}
\begin{center}

{\bf{\Large On the Possibility of Superluminal Neutrino Propagation}}

\vspace{1cm}

{\bf Jean Alexandre}$^{1}$, {\bf John Ellis}$^{1,2}$  and {\bf Nick E. Mavromatos}$^{1,2}$ \\
\vspace{0.5cm}
$^1$~Department of Physics, King's College London, Strand, London WC2R 2LS, UK \\
$^2$~Theory Division, Department of Physics, CERN, CH 1211 Geneva 23, Switzerland. 

\vspace{1cm}

{\bf Abstract}

\end{center}

We analyze the possibility of superluminal neutrino propagation $\delta v \equiv (v - c)/c > 0$ as indicated by OPERA data, 
in view of previous phenomenological constraints from supernova SN1987a and gravitational {\v C}erenkov
radiation. We argue that the SN1987a data rule out $\delta v \sim (E_\nu/M_N)^N$ for $N \le 2$
and exclude, in particular, a Lorentz-invariant
interpretation in terms of a `conventional'  tachyonic neutrino. We present two toy Lorentz-violating
theoretical models, one a Lifshitz-type fermion model with superluminality depending quadratically
on energy, and the other a
Lorentz-violating modification of a massless Abelian gauge theory with axial-vector couplings to fermions.
In the presence of an appropriate background field, fermions may propagate superluminally or subluminally,
depending inversely on energy, and on direction.
Reconciling OPERA with SN1987a would require this background field to depend on location.

\vspace{1cm}

\section{Introduction}

Data from the OPERA experiment have recently been interpreted~\cite{opera} as evidence for superluminal $\nu_\mu$ propagation
between CERN and the Gran Sasso laboratory, with 
$\delta v \equiv (v - c)/c \sim 2.5 \times 10^{-5}$ for $\langle E_\nu \rangle \sim 28$~GeV~\footnote{OPERA
used a similar experimental approach to that proposed in~\cite{harries}.}.
Such an extraordinary claim clearly requires
extraordinary standards of proof, notably including confirmation by an independent experiment such
as MINOS, T2K or NO$\nu$A. Nevertheless, 
%the reported effect has a significance $ \sim 6 \sigma$, so should be treated with some respect, 
even while the OPERA data are undergoing
experimental scrutiny, notably of the technical issues of pulse modelling, timing and distance measurement
on which we are not qualified to comment, it may be helpful to present some relevant phenomenological
and theoretical observations about the claimed effect. Here we report two sets of
considerations concerning: (1) comparison with other phenomenological
constraints on possible superluminal neutrino propagation, and (2) instructive theoretical toy models
of Lorentz violation that exemplify the price to be paid to obtain such an effect. These toy models
cast light on possible experimental probes of the OPERA effect.

As we show, reconciling this effect with other bounds on the propagation speeds of neutrinos,
notably those provided by the supernova SN1987a~\cite{SN1987a,oldSN,harries}, is a non-trivial issue. For example, if
$\delta v$ were independent of energy, the SN1987a neutrinos would have arrived at Earth
years before their optical counterparts. This prematurity would have been even more pronounced
for `conventional' Lorentz-invariant tachyons, for which $\delta v$ would {\it increase}
at lower energies, forcing one to consider Lorentz-violating models.
However, simple Lorentz-violating power-law modifications of the neutrino propagation speed $\delta v \sim (E_\nu/M_N)^N$
are also severely constrained by SN1987a. Specifically, constraints for $N = 1, 2$,
derived previously in the paper~\cite{harries} of which one of us (J.E.) was an author, 
are incompatible with the OPERA result for $\langle E_\nu \rangle \sim 28$~GeV~\cite{opera}.
Moreover, OPERA reports~\cite{opera} that there is no significant difference between the
values of $\delta v$ measured for the lower- and higher-energy data with $\langle E_\nu \rangle
\sim 13$ and 43~GeV, respectively, providing no indication that $N \ne 0$.%, rendering implausible models with $N > 2$. 

We also discuss the constraints imposed on superluminal neutrino
propagation by {\v C}erenkov radiation {\it in vacuo}. Electromagnetic {\v C}erenkov radiation
%JEis not a big issue, as it 
is suppressed by the absence of an electric charge for the neutrino~\cite{nuCer}.
%and can arise only via the neutrino charge radius~\cite{nuradius}. 
However, {\it gravitational}
{\v C}erenkov radiation~\cite{MN} is potentially significant for high-energy neutrinos, and an effect
of the type reported by OPERA could suppress high-energy astrophysical neutrino signals such as
those associated with the GZK cutoff and with gamma-ray bursters.

As a complement to these phenomenological remarks, we present two models for
Lorentz-violating fermion propagation, with different energy dependences for the
superluminality. One is a simple renormalizable Lifshitz-type fermion model in which the
superluminality {\it increases quadratically} with energy. This model also exhibits
dynamical fermion mass generation and asymptotic freedom. The other is a modification of an earlier gauge model for
Lorentz violation with {\it sub}luminal neutrino propagation that proposed previously by two of us (J.A.
and N.E.M.)~\cite{almav}. The modified model has a background axial U(1) gauge field, and may exhibit
{\it super}luminal neutrino propagation with $\delta v$ {\it falling as the inverse of the energy}
and depending on direction.
Since superluminal propagation with $\delta v \sim 2.5 \times 10^{-5}$ or greater is incompatible
with the SN1987a data, such a model could be
compatible with the data only if the background field depends on spatial location,
and is enhanced in the neighbourhood of the Earth compared to its mean value
along the line of sight to the Larger Magellanic Cloud.

\section{Phenomenological Constraints on Superluminal Neutrino Propagation}

The primary OPERA result on the mean neutrino propagation speed is
\begin{equation}
\delta v \; = \; (2.48 \pm 0.28 \pm 0.30) \times 10^{-5}
\label{deltav}
\end{equation}
for $\nu_\mu$ with $\langle E_\nu \rangle = 28$~GeV, where the errors in (\ref{deltav}) are
statistical and systematic, respectively.
%~\footnote{The rotation of the Earth is not entirely negligible. Between neutrino emission at CERN and arrival at the OPERA experiment, the latter moves eastwards, elongating the baseline by $\sim 50$~cm. This has the effect of slightly delaying the calculated arrival time, opposite in sign to the claimed effect.}. 
The OPERA Collaboration also provides the following
supplementary information on the difference in mean arrival times of samples of higher- and 
lower-energy neutrinos with $\langle E_\nu \rangle = 13$ and 43~GeV, $\Delta t = 14.0 \pm 26.2$~ns,
which corresponds to
\begin{equation}
\Delta(\delta v) \; = \; (0.57 \pm 1.07) \times 10^{-5}
\label{Deltadeltav}
\end{equation}
for the difference $\Delta(\delta v)$ between the propagation speeds of these neutrino samples~\footnote{We
note that no other experiment has made such accurate velocity measurements for any other particles with Lorentz 
boosts as large as the OPERA neutrinos.}.

Constraints on possible deviations of the speed of $\nu_\mu$ propagation from the
velocity of light had been placed previously by the MINOS Collaboration~\cite{MINOS},
which found
\begin{equation}
\delta v \; = \; (5.1 \pm 2.9) \times 10^{-5}
\label{MINOS}
\end{equation}
for $\nu_\mu$ with a spectrum peaking at $E_\nu \sim 3$~GeV and a tail extending
above 100~GeV. The MINOS result (\ref{MINOS}) is not significant in itself, but is also
compatible with the OPERA results (\ref{deltav}, \ref{Deltadeltav}), as are earlier neutrino
results~\cite{early}.

However, more
stringent constraints on models of neutrino propagation are imposed by the SN1987a neutrinos~\cite{SN1987a}.
The observed neutrinos emitted by SN1987a had energies around three orders of magnitude
smaller than the OPERA neutrinos. A significant fraction of them were undoubtedly $\nu_\mu$,
and neutrino oscillation phenomenology severely constrains differences in the propagation
speeds of different neutrino flavours, so the OPERA results (\ref{deltav}, \ref{Deltadeltav})
may be confronted directly with the SN1987a data. Since the distance to SN1987a was
$\sim 50$~kpc, i.e., $\sim 170,000$~light-years, an energy-dependent $\delta v$ of the
magnitude (\ref{deltav}) would have caused the SN1987a neutrinos to have arrived over 4~years
before their photon counterparts, whereas the maximum tolerable advance is only a few hours,
corresponding to $\delta v \sim 2 \times 10^{-9}$~\cite{oldSN}.

The SN1987a data are orders of magnitude more problematic for `conventional'
tachyonic neutrinos. Assuming Lorentz invariance, these would have a dispersion
relation $E^2 = p^2 - \mu^2$, where $\mu^2 > 0$, and the corresponding deviation of
the propagation speed from the velocity of light would be $\delta v \sim \mu^2/2E^2$. Thus, 
as the energy increases, the speeds of such `conventional' Lorentz-invariant tachyonic 
neutrinos would {\it decrease} towards the velocity of light. Normalizing the effect to
the value (\ref{deltav}) of $\delta v$ reported by OPERA for their relatively high-energy neutrinos
would lead to an impossibly large effect for the SN1987a neutrinos. Moreover, the
magnitude of $\mu^2$ would be incompatible with limits on the $\nu_\mu$ mass from $\pi$ and $\mu$ decay and,
to the extent that oscillation experiments constrain the $\nu_e - \nu_\mu$ mass difference, also the
direct constraint on the $\nu_e$ mass~\cite{PDG}.

We are therefore led to consider the possibility of Lorentz violation.
Although the OPERA Collaboration sees (\ref{Deltadeltav})
no significant energy dependence of $\delta v$ when comparing
its lower- and higher-energy samples with $\langle E_\nu \rangle \sim 13$ and 43~GeV~\cite{opera},
the SN1987a data motivate us to look at the implications of an energy
dependence $\delta v \sim (E_\nu/M_N)^N$ with
$\delta v  \sim 2.48 \times 10^{-5}$ for $\langle E_\nu \rangle \sim 28$~GeV. 
Under this hypothesis, the OPERA data would correspond to
\begin{eqnarray}
M_1 & \sim & 1.1 \times 10^6~{\rm GeV} , \label{M1} \\
{\rm or} \; \; \; M_2 & \sim & 5.6 \times 10^3~{\rm GeV} , \label{M2}
\end{eqnarray}
for linear and quadratic energy dependences, respectively. However, 
stringent constraints on $M_1$ and $M_2$ have been imposed previously by observations of the neutrino
burst from SN1987a~\cite{SN1987a}, which would have been spread out by any energy-dependence
of $\delta v$. The following constraints on superluminal neutrino propagation were established
by a collaboration including one of the present authors (J.E.)~\cite{harries}~\footnote{The
prospective sensitivity of the OPERA experiment to possible superluminal
neutrino propagation was also estimated in~\cite{harries}, using a similar technique and with with results similar to those now
obtained (\ref{M1}, \ref{M2}) by OPERA.}:
\begin{eqnarray}
M_1 & \sim & 2.5 \times 10^{10}~{\rm GeV} , \label{SNM1} \\
{\rm or} \; \; \; M_2 & \sim & 4 \times 10^4~{\rm GeV} . \label{SNM2}
\end{eqnarray}
We recall that the supernova neutrino burst is expected to have contained large
fractions of $\nu_\mu$ and ${\bar \nu}_\mu$, so that these constraints apply
{\it a priori} to the $\nu_\mu$ used by OPERA in their measurement.

Comparing (\ref{M1}) with (\ref{SNM1}),
we infer that a linear dependence of $\delta v$
is not compatible simultaneously with the OPERA and SN1987a data.
The situation with a quadratic energy dependence, cf, (\ref{M2}) and (\ref{SNM2}),
is not in such stark contradiction with the SN1987a data, but the latter would prefer
a stronger energy dependence that would be even more difficult to reconcile with
the lack of any indication of energy dependence within the OPERA data (\ref{Deltadeltav}).
As for a possible constant $\delta v$, we recall that the OPERA
measurement of $\delta v \sim 2.5 \times 10^{-5}$ would have led to the
SN1987a neutrino signal being observed $\sim 4$~years before the
optical signal, whereas the observed advance of $< 3$~hours (which is
compatible with models of supernova explosions) would correspond to
$\delta v < 2 \times 10^{-9}$~\cite{oldSN}. We infer that only an energy dependence
of $\delta v \sim E^N$ with $N > 2$ could reconcile the OPERA and SN1987a data,
though this is unlikely to be compatible with the lack of a significant energy dependence
observed within the OPERA energy range (\ref{Deltadeltav}).

The possibility of gravitational {\v C}erenkov radiation has been studied in~\cite{MN}.
The case studied there was that of a particle propagating at the speed of light
emitting {\it subluminal} gravitational radiation, but the same analysis applies to
a {\it superluminal} particle emitting gravitational radiation travelling at the
speed of light. It was shown in~\cite{MN} that a particle would lose all its energy
within a time
\begin{equation}
t_{max} \; = \; \frac{M_P^2}{(n-1)^2 E^3} ,
\label{GA}
\end{equation}
where $M_P \sim 1.2 \times 10^{19}$~GeV is the Planck mass and $n$ is the refractive
index: $n = 1 - \delta v$ in our case. Setting $\delta v \sim 2.5 \times 10^{-5}$ as
suggested by OPERA, using (\ref{GA}) we find
\begin{equation}
t_{max} \; \sim \; \frac{2 \times 10^{8}}{[E_\nu ({\rm GeV})]^3} \; {\rm years} .
\label{tmax}
\end{equation}
We conclude that applying the OPERA result simple-mindedly would exclude by
many orders of magnitude the observation of GZK neutrinos~\cite{GZK}, which should have 
$E_\nu \sim 10^{10}$~GeV and propagate $\sim 10^8$~light-years~\footnote{As for possible GeV-range neutrinos
emitted by gamma-ray bursters with cosmological redshifts, (\ref{tmax}) shows that they would
lose their energy before reaching the Earth, in addition to arriving at very different times from their
optical counterparts if $\delta v$ is given by(\ref{deltav}).}. Alternatively,
neutrinos with $E_\nu \sim 2 \times 10^6$~GeV, the minimum for which the
IceCube experiment has so far published an upper limit on the flux~\cite{IceCube}, could not
travel more than $\sim 10^{-4}$~seconds, ample to explain their non-observation,
though this surely has a less radical explanation! Conversely, observation of
neutrinos violating the bound (\ref{tmax}) would invalidate the hypothesis of
a constant $\delta v$ with the magnitude suggested by OPERA.

\section{Lorentz-Violating Models with Superluminal Fermion Propagation}

In light of the foregoing phenomenological discussion, one might be tempted to
lose interest in theories with superluminal neutrino propagation. However, the
effect reported by OPERA is so striking and of such potential significance that it
is important to study whether such an effect is possible, even in principle, and
how theoretical possibilities could be constrained by future experiments.
In this Section, we show how to construct examples within the general
framework of field theories with higher-order spatial derivatives, and
discuss some characteristic experimental signatures.

\subsection{Lifshitz-type Field Theory}

Such theories have recently attracted renewed attention because of their improved
convergence properties (for a recent review see~\cite{alexlif} and references therein).  In this spirit, a renormalizable Lifshitz-type theory of 
gravity has been proposed, which could lead to a renormalizable quantum gravity theory at 
high energies~\cite{Horava}. Such theories are free of ghosts, since the order of the time 
derivative in the action remains minimal, so that no new poles appear in particle propagators.
However, in general such theories violate Lorentz symmetry at 
high energies~\cite{kostelecky}.

We now exhibit a Lifshitz-type model with superluminal fermion propagation. The model
is formulated in three space dimensions with anisotropic scaling parameter $z=3$. In this scenario, 
the mass dimensions of the coordinates are  $[t]=-z = -3$, $[x]=-1$, and the free fermion action is~\cite{4ferm}
\be\label{free}
S_{4ferm}=\int dt d\vec x \left( \overline \psi i\gamma_0\dot\psi-\overline \psi(M^2-\Delta)(i\vec\partial\cdot\vec\gamma)\psi
+g(\ovl\psi\psi)^2\right)~,
\ee
where $\Delta \equiv -\partial_i\partial^i=\vec\partial\cdot\vec\partial$, and we use (+1,-1,-1,-1) as the metric signature.
The model is renormalizable since $[g]=0$, and we have also $[M]=1$, $[\psi]=3/2$. This model exhibits
asymptotic freedom as well as the dynamical generation of a mass $m_{dyn}$ for the fermion, as discussed in more
detail in~\cite{4ferm},\cite{alexlif}.

Taking this fermion dynamical mass into account, we obtain the following dispersion relation:
\be
\omega^2=m_{dyn}^6+M^4p^2+2M^2p^4+p^6~,
\label{model1}
\ee
and, assuming that $M\ne0$, it is possible to recover approximately Lorentz-invariant kinematics in the
infra-red limit, since the rescaling $\omega=M^2\tilde\omega$ leads to
\be\label{disp}
\tilde\omega^2=\mu_{dyn}^2+p^2+\frac{2}{M^2}p^4+\frac{p^6}{M^4},
\ee
where $\mu_{dyn} \equiv m_{dyn}^3/M^2$. Using this dispersion relation, one can compute the group velocity
$\frac{\partial {\tilde \omega}}{\partial p}$ and the phase velocity $\frac{\tilde\omega}{p}$ as power series in 
$(p/M)^{2N} p$: $N \ge 1$. The superluminal character of both follows immediately from (\ref{disp}), with the first 
correction of order $p^2/M^2$, which is quadratically suppressed by the Lorentz-violating mass scale $M$. 
We note that the \emph{superluminality} of this model is an unavoidable consequence of the relative signs of the 
various terms appearing in (\ref{free}), if one is to avoid tachyonic modes for sufficiently high momenta $p$.

It is clear from (\ref{model1}) that the superluminality $\delta v$ {\it increases} quadratically
with the fermion momentum (or energy) for $p (E) < M$, and even faster at higher 
momenta (energies). As discussed in the previous Section, such a quadratic dependence
is not easy to reconcile with the lack of energy dependence in $\delta v$ seen in OPERA data~\cite{opera},
though it comes closer to compatibility between the OPERA data and the constraint imposed
by SN1987a~\cite{harries}.

The fact that the superluminality in this model is quadratic in $E_\nu$ implies that {\it no effect}
should be seen at the level (\ref{deltav}) in the MINOS and T2K experiments, since they have
mean energies that are almost an order of magnitude lower than the CNGS beam. In particular, the
indication that $\delta v \ne 0$ from previous MINOS data~\cite{MINOS} would not be confirmed in this scenario.

\subsection{Lorentz-Violating Gauge Theory} 

We now consider more complicated models that lead to forms for $\delta v$ with very different
energy dependences, involving a fermion
coupling to either a vector or an axial U(1) gauge field. If there is a background field with a suitable constant value in a given 
reference frame,
these models may exhibit superluminal fermions, as well as other dramatic signatures highlighted below. 
These models are concrete realizations of the ideas
of~\cite{coleman}, where the phenomenology of Lorentz violation has been discussed in models where the maximal 
speeds for various particles depend on the species.

A minimal Lorentz-violating (LV) extension of massless Quantum Electrodynamics (QED) 
was proposed in~\cite{mdyn}, in which higher-order spatial derivatives were introduced 
for the photon field, and fermions remained minimally coupled to the photon. This theory
has the features that the light-cone `seen' by fermions differs from that `seen' by the photon.
Specifically, in the theory of~\cite{mdyn} (i) the photon always travels at the conventional speed of
light, (ii) fermions travel {\it subluminally}, and (iii) fermion masses may be generated
dynamically in such a framework, as an alternative to the Higgs mechanism.
We will show that similar theories with a background vector or axial U(1) field (see also~\cite{almav}) may lead to
{\it superluminal} fermion propagation, albeit with no mechanism for fermion mass
generation.

The Lagrangians of the models read:
\begin{equation}\label{chiral}
{\cal L}_{V,A} = -\frac{1}{4} G_{\mu\nu} \left( 1-\frac{\Delta}{M^2}\right) G^{\mu\nu}
+\ovl\psi \left( i\br\partial-g_{V,A} \br B \Gamma \underline{\tau} \right) \psi-m\overline\psi\psi~,
\end{equation}
where $G_{\mu\nu} \equiv \partial_\mu B_\nu - \partial_\nu B_\mu$ and $B_\mu$ is a gauge field with either a
vector coupling $g_V$ or an axial coupling $g_A$, depending whether $\Gamma = 1$ or $\gamma_5$,
respectively.
The presence of an axial $\gamma_5 \gamma^\mu $ fermion/gauge boson vertex would introduce the possibility of
chiral anomalies, which could be cancelled by suitable choices of the couplings to the different fermion fields 
$\psi = (\psi_1,\cdots,\psi_n)$, represented here by the matrix $\underline{\tau}$ with the property:
\be\label{cond}
{\rm tr}\{\underline\tau\} = 0~.
\ee
In the case of a doublet of fermions, for definiteness, one could use
\be\label{tau3}
\psi = \left( \begin{array}{c} \psi_1 \\  \psi_2 \end{array}\right) ~, \qquad \underline{\tau} \equiv \frac{1}{\sqrt{2}}\, \underline{\tau}_3 
\, = \, \frac{1}{\sqrt{2}}\, \left( \begin{array}{cc} 1 & 0 \\ 0 & -1 \end{array}\right) ~, \quad {\rm tr}(\underline{\tau}^2)=1 ~,
\ee
although other choices can be made~\cite{massivevector}, 
as long as the anomaly-free condition (\ref{cond}) is satisfied.
It should be noted that {\it no} higher-order spatial derivatives are introduced for the fermion
fields because, in order to respect gauge invariance, such terms would need to be of the form
\be
\frac{1}{M^{n-1}}\ovl\psi (i\vec D_5\cdot\vec\gamma)^n\psi~~~~~~n\ge 2,
\ee
whereby $D_5$ denotes the axial-gauge-field fermion covariant derivative, 
which would introrduce new, non-renormalizable couplings.
The Lorentz-violating modification proposed in the Lagrangian (\ref{chiral}) does \emph{not }
alter the photon dispersion relation, which remains relativistic, but {\it does} modify the fermion
propagator, as we discuss below. 

It was observed in~\cite{branes} that models of this type can be obtained by considering the
propagation of photons and charged fermions in a D-particle model of space-time foam~\cite{Dfoam}, 
according to which our world is viewed as a 3-brane propagating in a higher-dimensional bulk space
that is punctured by point-like D0-brane defects (D-particles). Such models my lead to non-trivial optical 
properties of the vacuum, because electrically-neutral matter excitations, such as photons and neutrinos, 
may acquire non-trivial refractive indices through non-trivial interactions with the D-foam. 
In previous D-foam models~\cite{Dfoam}, these interactions led to \emph{subluminal} propagation. 
In the flat space-time limit, where the low-energy Lagrangian is derived, the microscopic reason why 
fermions do not have higher-derivative modifications was that charge conservation forbids interactions of
charged fermions with the foam. 

The explicit Lorentz violation due to the higher-spatial-derivative 
term in the action (\ref{chiral}) implies that the light-cone `seen' by the fermions is different from that `seen' by the
gauge boson, assumed here to be (almost) massless like the photon, always travels at $c$, the speed of light {\it in vacuo}. Specifically,
the maximal speed for the fermions is \emph{smaller} than $c$, as in the vector models of refs.~\cite{mdyn,almav}. 
This may be seen by following the one-loop analysis of the fermion wave-function renormalization
calculated in~\cite{mdyn}. Due to the higher-order spatial derivatives in (\ref{chiral}), the 
one-loop quantum corrections to the fermion kinetic terms are different for time and
space derivatives. 
A similar computation as the one made in \cite{mdyn} yields corrections of the form
\be
i\ovl\psi \left((1+Z_0)\partial_0\gamma^0-(1+Z_1)\vec\partial\cdot\vec\gamma\right) \psi~, 
\ee
where
\bea
Z_0&=& -\frac{2\alpha_{V,A}}{\pi}\left(\frac{1}{4}\ln(1/\mu) +\ln2 -\frac{1}{2}\right)+{\cal O}(\mu^2\ln(1/\mu)) , \nn
Z_1&=& -\frac{2\alpha_{V,A}}{\pi}\left(\frac{1}{4}\ln(1/\mu)+\frac{25}{18}-\frac{5}{3}\ln2\right)+{\cal O}(\mu^2\ln(1/\mu))~.
\label{Z0Z1}
\eea
where $\alpha_{V,A} \equiv g_{V,A}^2/4 \pi$ and $\mu \equiv m/M$. 
We note that the dominant terms in (\ref{Z0Z1}), which are proportional to $\ln(1/\mu)$, are the same for $Z_0$ and $Z_1$.
This is to be expected since, in the Lorentz-invariant limit: $M\to\infty$ and hence $\mu \to 0$
for fixed fermion mass, we must have $Z_0=Z_1$. 
After redefinition of the bare parameters in the minimal substraction scheme, where
only the terms proportional to $\ln(1/\mu)$ are absorbed, the fermion dispersion relation is
\be\label{fdisp}
\left(1- \frac{\alpha_{V,A}}{\pi}[2\ln2-1]\right)^2\omega^2=\left(1- \frac{\alpha_{V,A}}{\pi}[25/9-(10/3)\ln2]\right)^2p^2+m^2~.
\ee
and the fermion phase and group velocities $v_\phi$, $v_g$ are both subluminal: 
\be\label{phase}
v_\phi =   v_g \; = \; 1 - \frac{\alpha_{V,A}}{\pi}\left(\frac{34}{9} - \frac{16}{3}\ln2\right) +{\cal O}(\alpha_{V,A}^2)~< 1~,
\ee
whereas the gauge boson $B_\mu$
propagates with the standard speed of light \emph{in vacuo}, $c$, as required by gauge invariance~\cite{mdyn}. 

A few remarks are in order at this point. First: in view of (\ref{phase}), the above models constitute
explicit microscopic realizations of the class of Lorentz-violating theories of the type considered in \cite{coleman}, 
with species-dependent light cones. 
Secondly, the fact that the constant wave function renormalization (\ref{Z0Z1}) is found to
be less than one, which leads to the subluminal velocities (\ref{phase}),
is a rather general property of quantum field theory, stemming from unitarity~\cite{higa}. 
Indeed, in field theories with non-negative-metric states, 
the wave function renormalization $A$ must satisfy $0 <  A  <  1$, which also implies non-negative anomalous dimensions. 
However, there may be cases, e.g., with derivative interactions~\cite{higa}, in which negative anomalous dimensions appear, 
with the consequence that the wavefunction renormalization can be larger than one. It would be interesting to investigate
the possibility of superluminal propagation in such cases, by analogy with the scenario discussed above.

We now explore the possibility of superluminal fermion propagation in the context of the above theories.
To this end, we first consider the possibility of a constant background gauge field $B_\mu^{(0)}$,
in which case the relevant part of the action (\ref{chiral}) reduces to:
\be\label{axial}
{\cal L}_{\rm bckgrd} =  \ovl\psi \left( i\br\partial-g_{V,A} \br B^{(0)} \Gamma \underline{\tau} \right) \psi-m\overline\psi\psi~.
\ee
 These models fall within the general category of Lorentz-violating extensions of the Standard Model~\cite{kostelecky},
as reviewed in the specific case of neutrinos in~\cite{diaz}, taking into account the available neutrino oscillation data. 
Depending on the sign of the background field $B^{(0)}$ in (\ref{axial}), one may have group velocities for the fermions which
are superluminal. The quantum fluctuations of the axial gauge field would tend to counteract such superluminality, 
as discussed above (\ref{phase}). Nevertheless, for sufficiently weak couplings $\alpha_{V,A}$ and a
sufficiently strong background field $B^{(0)}$, the maximal fermion speed may be superluminal.
A value of $\delta v \sim 2.5 \times 10^{-5}$, as reported by OPERA~\cite{opera}, 
may be arranged in these models with a small, perturbative gauge coupling $g_{V,A} < 1$ and a
background field $B^{(0)}$ of appropriate magnitude.

The vector (axial) interaction of  (\ref{axial}) has the same (different) 
signs for left- and right-handed fermions, such as neutrinos and their antiparticles $\psi^c$,
which we assume to be Majorana fermions. This could lead to a physically 
important difference between the dispersion relations of neutrinos and antineutrinos,
and hence apparent CPT violation:
\begin{eqnarray}
\omega_\nu & = & \sqrt{ (\vec{p} - g_{V,A} \vec{B})^2 + m^2} + g_{V,A} B_0 ~, \nonumber \\
\omega_{\overline \nu} & = & \sqrt{ (\vec{p} \mp g_{V,A} \vec{B})^2 + m^2} \pm g_{V,A} B_0~.
\label{dispnunubar}
\end{eqnarray}
where the upper (lower) symbols in the combinations $\pm, \mp$ refer to the vector (axial) case~\footnote{In the simple two-flavour 
axial model (\ref{tau3}), the particle of one flavour would exhibit the same dispersion relation as the antiparticle of the other flavour.}$^,$\footnote{An effect similar to the axial case in (\ref{dispnunubar}), but without the flavour structure, 
could arise purely geometrically in the propagation of fermions in space-times that break rotational symmetry, 
such as rotating Kerr black holes or axisymmetric Robertson-Walker Universes,  as discussed in~\cite{mukho}. 
Such geometric effects stem from the coupling of the spin of the fermions to non-trivial local curvature effects that 
arise in such space-times.}. Notice that these dispersion relations are the usual ones for massive particles, 
though with generalised momenta
\begin{equation} \label{Pi}
\Pi^0 = \omega_\nu \mp g_{A,V} B^0, \quad \vec{\Pi} = \vec{p} \mp g_{A,V} \vec{B} ,
\end{equation}
where the upper signs apply to neutrinos, and to antineutrinos with a vector interaction, and the lower signs apply
to antineutrinos with an axial interaction.

Assuming that the components $\vec{B}, B_0$ are constants in a local frame of reference, and
defining the angle between the three-vectors 
$\vec{p}$ and $\vec{B}$ to be $\vartheta$, we may write the phase velocity following from 
(\ref{dispnunubar}) for high-energy neutrinos with $p \gg m$ as:
\begin{equation}\label{groupph}
v_{ph} = \frac{\omega_\nu}{p} \simeq 1 \mp \frac{g_{V,A}}{p}(|\vec B|\cos\vartheta-B_0)+\cdots ~,
\end{equation}
where dots represent higher orders in $1/p$.
We obtain a similar expression for antineutrinos but with the replacement $|\vec B| \rightarrow -|\vec B|$ and 
$B_0\rightarrow -B_0$ in the axial case. 
However, the superluminality associated with (\ref{groupph}) 
does not apply to the group velocity,  which is subluminal:
\be
v_g = \frac{\partial \omega_\nu}{\partial p} = 1 - \frac{1}{2p^2}(g_{V,A}^2B^2\sin^2\vartheta + m^2)+\cdots~,
\ee
which is the same for neutrinos and antineutrinos, and
where the dots represent higher orders in $1/p$. Note that $|v_g-1|$ is of order $1/p^2$, unlike the case of the phase velocity, where
$|v_{ph}-1|$ is of order $1/p$~\footnote{ This model has the
interesting feature that the (anti)neutrino propagation velocity depends on the direction of propagation. 
This example raises the possibility that,
if a constant limiting velocity of light does not apply to neutrinos, perhaps the Michelson-Morley experiment should also
be revisited for neutrinos?}.

As a further step, we modify the background space-time in which the neutrino propagates, exhibiting
an extension of this model with superluminal
group velocities. We embed the model (\ref{axial}) in a modification of
Minkowskian space-time with non-diagonal metric components
that break the rotational symmetry along a specific axis, 
\begin{equation}\label{axisym}
g_{0i} = \vec{V}_i \,, \quad   i=1,2,3
\end{equation}
where $|\vec V|\equiv V  \ll 1$ is considered as a small perturbation~\footnote{This is motivated
by the suggestion that the metric distortion (\ref{axisym}) and the axial background case (\ref{axial}) may have a 
common geometric origin, given that they may both be associated with background space-time effects, 
with the vector $\vec{B}$ pertaining to the coupling of the (anti)neutrino spin to local curvature effects~\cite{mukho}, as
mentioned above.}. 
For {\it constant and homogeneous} $\vec V$, the dispersion relations (\ref{dispnunubar}) for neutrinos are modified to 
\be\label{disps}
\Pi ^\mu \Pi ^\nu   g_{\mu\nu} =  m^2~, 
\ee
where $\Pi^\mu$ is given by (\ref{Pi}). From this we obtain:
\begin{eqnarray}
\omega_\nu & = & - (\vec{p}-g_{V,A}\vec B) \cdot \vec{V} +  
\sqrt{(\vec{p} - g_{V,A} \vec{B})^2 + m^2} + g_{V,A} B_0 + O(V^2) ~, \nonumber \\
\omega_{\overline \nu} & = & - (\vec{p}-g_{V,A}\vec B) \cdot \vec{V}  +   
\sqrt{(\vec{p} \mp g_{V,A} \vec{B})^2 + m^2} \pm g_{V,A} B_0  + O(V^2)~.
\label{dispnunubar2}
\end{eqnarray}
Assuming that the components $\vec{B}, B_0$ are constants in a local frame of reference,
considering for simplicity the case with $|\vec{V}| \ll |\vec{B}|$, and defining the angle between the three-vectors 
$\vec{p}$ and $\vec{B}$ 
to be $\vartheta$, and that between $\vec{p}$ and $\vec{V}$ to be $\varphi$, then 
we observe that Eq.~ (\ref{dispnunubar2})
yields the following expressions for the neutrino phase and group velocities 
for relatively high-energies: $p=|\vec{p}| \gg m, |\vec B|$:
\begin{eqnarray}\label{group}
v_{ph} & = & 1 - V {\rm cos}\varphi + \frac{g_{V,A} \vec{B} \cdot \vec{V}}{p} - \frac{g_{V,A}}{p}(|\vec B|\cos\vartheta-B_0)+\cdots \nonumber \\
v_{g} & = & 1 -  V \, {\rm cos} \varphi  - \frac{g_{V,A}^2 B^2 {\rm sin}^2 \vartheta + m^2 }{2p^2} + O(V^2) ~,
\end{eqnarray}
and similarly for antineutrinos but with the replacement $B \rightarrow - B$ in the axial case. Superluminal group 
velocities of order $\delta v \sim 2.5 \times 10^{-5}$, as reported by
the OPERA experiment~\cite{opera}, could be obtained for suitable values of the combination 
$-V {\rm cos}\varphi > 0$.

The model (\ref{axial}, \ref{axisym}, \ref{group}) has several dramatic and testable consequences:

$\bullet$ The deviation of the neutrino propagation speed from that of light could exhibit non-trivial
dependence on $E_\nu$, due to the combination of terms in (\ref{group}), that is not a simple power law.
Thus, compatibility with the MINOS result~\cite{MINOS} is a non-trivial issue, which we address below.

$\bullet$ The neutrino group velocity would depend on the angle of propagation. This means that the speed
of propagation would, in general, vary sinusoidally during the sidereal day, and could even vary between super-
and subluminality~\footnote{However, we would not expect any day-night or seasonal dependence, which is consistent
with the absences of such effects in the OPERA data~\cite{opera}.}.
This modulation would be absent only for $\vec{V}$ oriented parallel to the Earth's rotational axis.

$\bullet$ The amount of superluminality would also, in general, depend on the geographical orientation of the neutrino
beam. For example, in the hypothetical example in which $\vec{V}$ is oriented parallel to the Earth's rotational axis,
the sign of the effect on neutrinos travelling northwards (cf, the Fermilab-Soudan neutrino beam) would be
opposite to beams travelling in a southerly direction (cf, the CNGS neutrino beam), and would be almost null for a
beam oriented almost east-west (cf, the T2K neutrino beam). Studying the compatibility of MINOS data~\cite{MINOS}
with this model must therefore take into account the ambiguity in the orientation of $\vec{V}$, as well as the energy
dependence of the superluminal effect in this model.

$\bullet$ It is possible that the orientation and magnitude of $\vec{V}$ and $\vec{B}$ vary on an interstellar scale, in which case
the SN1987a constraint on the neutrino velocity applies only to an average over space and time of the
possible superluminality effect, and there is no {\it a priori} contradiction with the OPERA result. 

$\bullet$ If the neutrino group velocity is superluminal, the corresponding \emph{antineutrino} group velocity
in the same direction would also be \emph{superluminal} in both the axial and vector cases.

Another possibility is that the vector $\vec{V}$ (\ref{axisym}) may be associated with distortions of space-time 
due to the interaction of the neutrino with space-time defects, as in stringy D-particle models of space-time foam~\cite{Dfoam},  
in which the vector $\vec{V}$ is associated with the average transfer of momentum from the neutrino to space-time defects 
with which it interacts during its propagation. In such a case, the metric would be of Finsler type, i.e., depending not only
on the space-time coordinates but also on momenta. This possibility is included within our formalism, but we do not
pursue it further here. We note, however, that in such models electric charge conservation (which is enforced by gauge
invariance) prevents
charged matter (such as electrons) from interacting non-trivially with the D-particle foam~\cite{Dfoam,branes}, 
so that only neutral excitations (such as photons and neutrinos) may be affected by the foam. This may provide a 
microscopic explanation of the fact that for electrons no deviations from special relativity have been observed
with a precision $\sim 10^{-9}$~\cite{nuCer}.

\section{Summary and Prospects} 

The report from OPERA of superluminal neutrino propagation is very surprising,
and it may well not survive further scrutiny. Moreover, as we have shown in the earlier part of this
paper, it is subject to constraints from studies of lower-energy neutrinos, specifically those emitted
by SN1987a~\cite{harries}, and would have implications for higher-energy astrophysical neutrinos. 
In particular, we have argued that the SN1987a data exclude a `conventional' Lorentz-invariant tachyonic neutrino 
interpretation of the OPERA data. On the other hand, as we have shown through
the toy models presented in the latter part of this paper, it is possible to construct Lorentz-violating
theories in which neutrinos travel faster than photons, which always travel at $c$. We
have exhibited such models in which the superluminality either increases or decreases with energy. 
Superluminal neutrinos should not be discarded as a phenomenological impossibility, but rather regarded as a
scenario to be probed and constrained by experiment. In particular, we have shown that
the effect could depend on the orientation of the neutrino beam%, and could be of opposite sign for antineutrinos
. For the moment, the OPERA
measurement provides a stimulus for investigating such scenarios, but Lorentz-violating
superluminal fermion propagation should not necessarily be discarded out of hand, even if the OPERA
result were not to be confirmed. 

\section*{Notes added}

A number of papers reacting to the OPERA effect~\cite{opera} appeared before ours~\cite{others}.
There is some overlap with the phenomenological considerations presented in~\cite{harries} and here,
but the models discussed here do not seem to have been discussed yet in this context. 

We also note that,
among the extensive literature since our paper was released, it has been pointed out~\cite{CG} 
that the modified {\v C}erenkov radiation process $\nu \to \nu e^+ e^-$ is
potentially an important mechanism for energy loss by superluminal neutrinos. A first direct experimental limit on this process and
on the distortion of the neutrino energy spectrum that it might induce has been reported~\cite{ICARUS}. We limit ourselves here to
noting that the rate for this process is very sensitive to the magnitude of $\delta v$, and also to its energy dependence. 
We leave for future work a detailed combined study of the interplay between this and other constraints, pending verification of
the magnitude of the OPERA result and its energy dependence.
. 

In this context, we also note that in any model with general coordinate invariance, such as our model (\ref{axial}), 
where the modified dispersion relations (\ref{disps}) arise as a result of a non-trivial metric background,
e.g., (\ref{axisym}), one may always find coordinate transformations to a frame in which  the superluminal effects are absent.
For the background (\ref{axisym}), responsible for the superluminal $V$-dependent parts of the group velocity (\ref{group}), 
such transformations are of the Galilean form $t \to  t , \, x^i  \to x^i - V^i t$ , which, from the point of view of
a passive observer, result in a change in the metric $\delta g_{0i} = -V_i $ that can cancel the superluminal effects in the 
dispersion relation in that frame. Since the {\v C}erenkov radiation is a physical (observer-independent) phenomenon, 
it cannot depend on the coordinate choice made by the observer, whereas the refractive index can, being frame-dependent. 
Hence, we conclude that the arguments of ref.~\cite{CG} do not apply directly to our second model. We note that this argument 
would imply that, in the transformed frame, the dispersion relations of other particles, such as electrons, are affected. However, 
this is not in contradiction with the current bounds for these particles, which are derived in different experimental conditions,
and specifically in a different reference frame.

\section*{Acknowledgments} 

This work of J.E. and N.E.M. was supported in part by the London Centre for
Terauniverse Studies (LCTS), using funding from the European Research
Council via the Advanced Investigator Grant 267352.

  \end{document}